\documentclass[traditabstract]{aa}

\usepackage{graphicx}
\usepackage{txfonts}

\usepackage{natbib}
\bibpunct{(}{)}{;}{a}{}{,}

\def\nul#1{}
\def\ex{\rm e}
\def\be\begin{equation}
\def\ee\end{equation}
\def\ba\begin{array}
\def\ea\end{array}
\def\figpath{}

\begin{document}

\title{The HARPS search for southern extra-solar planets\thanks{Based
    on observations made with the HARPS instrument on the ESO 3.6 m
    telescope at La Silla Observatory under the GTO programme ID
    072.C-0488; and on observations obtained at the Keck Observatory,
    which is operated jointly by the University of California and the
    California Institute of Technology. The table with the HARPS radial
    velocities is available in electronic form at the CDS via anonymous ftp to
    cdsarc.u-strasbg.fr (130.79.128.5) or via http://cdsweb.u-strasbg.fr/cgi-bin/qcat?J/A+A/????  
}  }
   \subtitle{XIX. Characterization and dynamics of the {GJ}\,876 planetary system}
   \titlerunning{Characterization and dynamics of the planets in {GJ}\,876.}

\author{
A.C.M.~Correia\inst{1,2}
\and J.~Couetdic\inst{2}
\and J.~Laskar\inst{2}
\and X.~Bonfils\inst{3}
\and M.~Mayor\inst{4}
\and J.-L.~Bertaux\inst{5}
\and F.~Bouchy\inst{6}
\and X.~Delfosse\inst{3}
\and T.~Forveille\inst{3}
\and C.~Lovis\inst{4}
\and F.~Pepe\inst{4}
\and C.~Perrier\inst{3}
\and D.~Queloz\inst{4}
\and S.~Udry\inst{4}
}

%\offprints{}
 
\institute{Departamento de F\'isica, Universidade de Aveiro, Campus de
  Santiago, 3810-193 Aveiro, Portugal
  \and IMCCE, CNRS-UMR8028, Observatoire de Paris, UPMC, 77 avenue
  Denfert-Rochereau, 75014 Paris, France 
  \and Laboratoire d'Astrophysique, Observatoire de Grenoble, Universit\'e J.
  Fourier, CNRS-UMR5571, BP 53, 38041 Grenoble, France 
  \and Observatoire de Gen\`eve, Universit\'e de Gen\`eve, 51 ch.
  des Maillettes, 1290 Sauverny,  Switzerland
  \and Service d'A\'eronomie du CNRS/IPSL, Universit\'e de Versailles
  Saint-Quentin, BP3, 91371 Verri\`eres-le-Buisson, France 
  \and Institut d'Astrophysique de Paris, CNRS, Universit\'e Pierre et
  Marie Curie, 98bis Bd Arago, 75014 Paris, France 
}

\date{Received ; accepted To be inserted later}

  \abstract{
  Precise radial-velocity measurements for data acquired with the {\small HARPS} spectrograph
  infer that three planets orbit the M4 dwarf star {GJ}\,876. In particular, we 
  confirm the existence of planet $d$, which orbits every 1.93785~days. We find that
  its orbit may have significant eccentricity (e=0.14), and %revise its minimum mass to
  deduce a more accurate estimate of its minimum mass of
  $ 6.3\,M_\oplus$. Dynamical modeling of the {\small HARPS} measurements combined
  with literature velocities from the Keck Observatory strongly constrain the
  orbital inclinations of the $b$ and $c$  planets. We find that $i_b = 48.9^\circ \pm
  1.0^\circ $ and $i_c = 48.1^\circ \pm 2.1^\circ$, which infers the true
  planet masses of $M_b = 2.64 \pm 0.04\,M_{\rm Jup}$ and $ M_c = 0.83 \pm
  0.03\,M_{\rm Jup}$, respectively. 
  Radial velocities alone, in this favorable case, can therefore fully determine
  the orbital architecture of a multi-planet system, without the input from  
  astrometry or transits.
  
  The orbits of the two giant planets are nearly coplanar, 
  and their 2:1 mean motion resonance ensures stability over at least 5~Gyr. 
  The libration amplitude is smaller than $2^\circ$, suggesting that it was
  damped by some dissipative process during planet formation.
  The system has space for a stable fourth planet in a 4:1
  mean motion resonance with planet $b$, with a period around 15 days.
  The  radial velocity measurements constrain  the mass of this possible
  additional planet to be at most that of the Earth.
  } 


   \maketitle
%
%________________________________________________________________

\section{Introduction}

Also known as IL Aqr, GJ\,876 is only 4.72~pc away from our Sun and  is thus the 
closest star known to harbor a multi-planet system.  
It has been tracked by several instruments and telescopes from the very
beginning of the planetary hunt in 1994, namely the Hamilton (Lick),
HIRES (Keck) echelle spectrometers, ELODIE (Haute-Provence), and
CORALIE (La Silla) spectrographs. 
More recently, it was also included in the HARPS program.

The {\it {\small HARPS} search for southern extra-solar planets} is an extensive 
radial-velocity survey of some 2000 stars in the solar neighborhood conducted with the HARPS 
spectrograph on the ESO 3.6-m telescope at La\,Silla (Chile) in the framework of Guaranteed 
Time Observations granted to the HARPS building consortium \citep{Mayor_etal_2003}.
About 10\% of the HARPS GTO time were dedicated to observe a
volume-limited sample of  $\sim$110 M dwarfs. This program has proven to be
very efficient in finding Neptunes
\citep{Bonfils_etal_2005b,Bonfils_etal_2007} and Super-Earths
\citep{Udry_etal_2007,Forveille_etal_2009,Mayor_etal_2009b} down to $m \sin i
=1.9~M_\oplus$. Because M dwarfs are more
favorable targets to searches for lower mass and/or cooler planets than around
Sun-like stars, the observational effort dedicated by ESO has
increased by a factor of four, to an extended sample of 300 stars.

The first planet around GJ\,876, a Jupiter-mass planet with a period of about 
61~days, was simultaneous reported by \citet{Delfosse_etal_1998} and 
\citet{Marcy_etal_1998}. Later, \citet{Marcy_etal_2001} found that the 
61-day signal was produced by two planets in a 2:1 
mean motion resonance, the inner one with 30-day period also being a gas giant.
While investigating the dynamical interactions between those two planets,
\citet{Rivera_etal_2005} found evidence of a third planet, with an orbital 
period less than 2~days and a minimum mass of 5.9\,$M_\oplus$, which at that
time was the lowest mass detected companion to a main-sequence star other than
the Sun.

Systems with two or more interacting planets dramatically improve our
ability to understand planetary formation and 
evolution, since dynamical analysis can both constrain their evolutionary 
history and more accurately determine their orbital  ``structure''.
Amongst known multi-planet systems, planets GJ\,876\,$b$ and $c$ 
have by far the strongest mutual gravitational interactions, and
all their orbital quantities change quite rapidly. Since the radial velocity 
of GJ\,876 has been monitored for the past 15 years, 
the true masses of these two
planets can be determined by adjusting the inclination of their orbital 
planes when fitting dynamical orbits to the observational data.
This was first attempted by \citet{Rivera_Lissauer_2001}, who found 
a broad minimum in the residuals to the observed radial velocities for
inclinations higher than $30^\circ$, the best-fit function corresponding to $\sim 37^\circ$. 
Astrometric observations of GJ\,876 with the {\it Hubble Space Telescope}
subsequently suggested a higher value of $\sim 85^\circ$
\citep{Benedict_etal_2002}. 
\citet{Rivera_etal_2005} re-examined the dynamical interactions
with many additional Keck radial velocity measurements, and found
an intermediate inclination of $\sim 50^\circ$. 
\citet{Bean_Seifahrt_2009} reconciled the astrometry and
radial velocities by performing a joint adjustment to the Keck and HST
datasets, and showed that both are consistent with $\sim 50^\circ$. 
Early inclination determinations were, 
in retrospect, affected by small-number statistics (for astrometry) and by 
a modest signal-to-noise ratio in the radial velocity residuals. %from a 2-Keplerian orbit.

In the present study, we reanalyze the GJ\,876 system, including 52 additional
high precision radial velocity measurements taken with the HARPS spectrograph.
We confirm the presence of a small planet $d$ and determine the
masses of $b$ and $c$ with great precision.
Section\,2 summarizes  information about the GJ\,876 star. Its derived
orbital solution is described in Sect.\,3, and the dynamical analysis of the
system is discussed in Sect.\,4. Finally, conclusions  
are drawn in Sect.\,5.

%________________________________________________________________

\section{Stellar characteristics of GJ\,876}

\object{GJ\,876} (\object{IL\,Aquarii}, \object{Ross\,780}, \object{HIP\,113020})
is a M4 dwarf \citep{Hawley_etal_1996} in the Aquarius constellation.
At $4.7~\mathrm{pc}$
\citep[$\pi = 212.69 \pm 2.10~\mathrm{mas}$ --][]{ESA_1997},
it is the 41$^\mathrm{st}$ closest known stellar
system\footnote{http://www.chara.gsu.edu/RECONS/TOP100.posted.htm} and  
only the 3$^\mathrm{rd}$  closest known planetary system (after
$\epsilon$ Eridani, and slightly further away than
\object{GJ 674}).\  %Gl674 @ 4.5 pc

Its photometry \citep[$V =  10.162 \pm 0.009$,
$K = 5.010 \pm 0.021$ --][]{Turon_etal_1993, Cutri_etal_2003} and its
parallax imply absolute magnitudes of
$M_V = 11.80 \pm 0.05$ and $M_K = 6.65 \pm 0.05$.
The $J-K$ color of \object{GJ 876}
\citep[$=0.92$ --][]{Cutri_etal_2003} and
the \citet{Leggett_etal_2001} color-bolometric relation infer
a K-band bolometric correction of $BC_K=2.80$, and
a 0.013 $L_\odot$ luminosity.
The K-band mass-luminosity relation of
\cite{Delfosse_etal_2000} implies a $0.334 M_\odot$
mass, which is comparable to the $0.32 M_\odot$ derived by
\citet{Rivera_etal_2005} from \citet{Henry_McCarthy_1993}. 
\citet{Bonfils_etal_2005a} estimate it has an
approximately solar metallicity ($[\mathrm{Fe/H}]=0.02\pm0.2$). 
\citet{Johnson_Apps_2009} revised this metallicity calibration for the
most metal-rich M dwarfs and found that GJ876 has an above solar metallicity
($\mathrm{[Fe/H]}>+0.3 \mathrm{dex}$).

In terms of magnetic activity, \object{GJ 876} is also an average
star in our sample regarding its . 
Its \ion{Ca}{ii} H\&K emission is almost twice the emission of Barnard's star, an old
star in our sample of the same spectral type, but still comparable
to many stars with low jitter (rms$\sim2~\mathrm{m\,s^{-1}}$). On the
one hand, its long rotational period and magnetic activity imply an
old age ($>0.1$~Gyr). On the other hand, its UVW Galactic velocities
place \object{GJ~876} in the young disk population
\citep{Leggett_1992}, suggesting an age $<5$ Gyr, and therefore
bracketing its age to $\sim0.1 - 5$~Gyr.

\citet{Rivera_etal_2005} monitored \object{GJ 876} for photometric
variability and found a 96.7-day periodic variation with a $\sim$1\%
amplitude, hence identifying the stellar rotation. This short
rotational period is particularly helpful for radial-velocity
measurements as a dark spot covering 1\% of \object{GJ 876}'s surface
and located on its equator would produce a Doppler modulation with a
maximum semi-amplitude of only $\sim$1.5 m/s.

Finally, the high proper motion of
\object{GJ~876}
\citep[$1.17~\mathrm{arcsec\,yr^{-1}}$ --][]{ESA_1997}
changes the orientation of its velocity vector
along the line of sight \citep[e.g.][]{Kurster_etal_2003}
resulting in an apparent secular acceleration of
$0.15 \,\mathrm{m\,s^{-1}yr^{-1}}$.  Before our
orbital analysis, we removed this drift for both HARPS and already published data.

 \begin{table}
 \caption{Observed and inferred stellar parameters of GJ\,876.
 \label{T1} }
 \begin{center}
 \begin{tabular}{l l c c}
 \hline\hline
 \multicolumn{2}{l}{\bf Parameter} &\hspace*{2mm} & \bf GJ\,876 \\
 \hline
 Sp & & & M4\,V  \\
 $V$ & [mag] & & 10.162 \\
 $\pi$ & [mas] & & 212.69 $\pm$ 2.10 \\
 $M_V$ & [mag] & & 11.80 $\pm$ 0.05 \\
 $\mathrm{[Fe/H]}$ & [dex] & & $0.05$ $\pm$ 0.20 \\
 $L$ & [$L_{\odot}$] & & 0.013 \\
 $M_*$ & [$M_{\odot}$] & & 0.334 $\pm$ 10\% \\
 $P_{\mathrm{rot}}$& [day] & & 96.7 \\
 $v\sin{i}$ & [km s$^{-1}$] & & $0.16$ \\
 age($\log R'_{\mathrm{HK}}$) & [Gyr] & & 0.1$-$5 \\
 \hline
 \end{tabular}
 \end{center}
\end{table}

%________________________________________________________________

\section{Orbital solution for the GJ\,876 system}
\label{orbsolutions}

\subsection{Keck + HARPS}

The {HARPS} observations of {GJ}\,876 started in December 2003 
and have continued for more than four years. 
During this time, we acquired 52 radial-velocity measurements with an
average precision of 1.0~m/s.

Before the {HARPS} program, the star GJ\,876 had already
been followed since June 1997 by the {HIRES} spectrometer 
mounted on the 10-m telescope I at Keck Observatory (Hawaii).
The last published set of data acquired at Keck are from 
December 2004 and correspond to 155 radial-velocity measurements 
with an average precision of 4.3~m/s \citep{Rivera_etal_2005}.

When combining the Keck and the HARPS data for {GJ}\,876, the time span %4102.6
for the observations increases to more than 11 years, and the secular dynamics of
the system can be far more tightly constrained, although the average precision
of the Keck measurements is about four times less accurate than for HARPS.

\subsection{Two planet solution}

With 207 measurements (155 from Keck and 52 from HARPS), 
we are now able to determine the nature of the three bodies in the system with
great accuracy.
Using the iterative Levenberg-Marquardt method \citep{Press_etal_1992},
we first attempt to fit the complete set of radial
velocities with a 3-body Newtonian model (two planets)
assuming coplanar motion perpendicular to the plane of the sky, similarly to
that achieved for the system {\small HD}\,202206 \citep{Correia_etal_2005}.
This fit yields the well-known planetary companions at $P_b=61$~day, and
$P_c=30$~day, and an adjustment of $\sqrt{\chi^2}=2.64$ and
$rms=4.52\mathrm{m/s}$. 

\subsection{Fitting the inclinations}

Because of the proximity of the two planets and their high minimum masses, the
gravitational interactions between these two bodies are strong.  
Since the observational data cover more than one decade, we can expect to
constrain the inclination of their orbital planes. 
Without loss of
generality, we use the plane of the sky as a reference plane,
and choose the line of nodes of planet $b$ as a reference
direction in this plane ($\Omega_b = 0$).  
We thus add only three   free
parameters to our fit: the longitude of the node $\Omega_c$ of
planet $c$,  and the inclinations  $i_b$ and $i_c$ 
of planets $b$ and $c$ with respect to
the plane of the sky.

\begin{figure}
   \centering
    \includegraphics*[height=8.5cm,angle=270]{\figpath 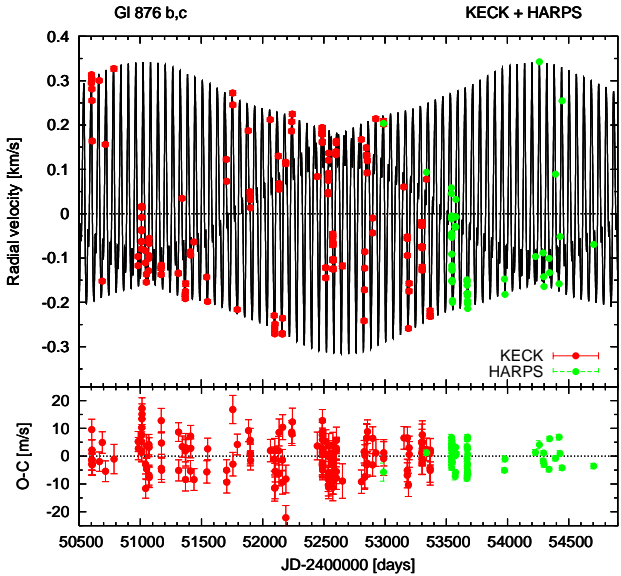}
  \caption{Keck and HARPS radial velocities for {GJ}\,876, superimposed on the
  3-body Newtonian (two planets) orbital solution for planets $b$ and $c$ (orbital parameters
  taken from Table\,\ref{T2}). 
    \label{F1} }
\end{figure}

This fit yields $i_b=57^\circ$, $i_c=63^\circ$, and
$\Omega_c=-1^\circ$, and an adjustment of $\sqrt{\chi^2}=2.57$ and
$rms=4.35\mathrm{m/s}$. 
Although there is no significant improvement to the fit, an important difference
exists: the new orbital parameters for both planets show some 
deviations and, more importantly, the inclinations of both planets are well
constrained.
In Fig.~\ref{F1}, we plot the Keck and HARPS radial velocities for {GJ}\,876,
superimposed on the 3-body Newtonian orbital solution for planets $b$ and $c$.

\subsection{Third planet solution}

\begin{table*}
\caption{Orbital parameters for the planets orbiting {GJ}\,876, obtained
with a 4-body Newtonian fit to observational data from Keck and HARPS.  \label{T2} }
\begin{center}
\begin{tabular}{l l c c c} \hline \hline
\noalign{\smallskip}
{\bf Param.}  & {\bf [unit]} & {\bf b} & {\bf c} & {\bf d} \\ \hline 
\noalign{\smallskip}
Date         & [JD]             & \multicolumn{3}{c}{2\,455\,000.00 (fixed)}  \\ 
$V_{(Keck)}$          & [km/s]    & \multicolumn{3}{c}{$\phantom{-}0.0130 \pm 0.0004 $}  \\ 
$V_{(HARPS)}$         & [km/s]    & \multicolumn{3}{c}{$-1.3388 \pm 0.0004 $}  \\ \hline  
\noalign{\smallskip}
$P$          & [day]                & $ 61.067 \pm 0.011 $ & $ 30.258 \pm 0.009 $ & $ 1.93785 \pm 0.00002  $ \\ 
$\lambda$    & [deg]                & $ 35.61  \pm 0.14  $ & $ 158.62 \pm 0.80  $ & $   29.94 \pm 3.30  $ \\ 
$e$          &                      & $  0.029 \pm 0.001 $ & $  0.266 \pm 0.003 $ & $   0.139 \pm 0.032 $ \\ 
$\omega$     & [deg]                & $ 275.52 \pm 2.67  $ & $ 275.26 \pm 1.25  $ & $  170.60 \pm 15.52 $ \\ 
$K$          & [m/s]                & $ 212.24 \pm 0.33  $ & $  86.15 \pm 0.40  $ & $    6.67 \pm 0.26  $ \\  
%$T$          & [JD-2400000]         & $ 54979.6 \pm 0.5 $ & $ 54979.4 \pm 0.1 $ & $ 54998.8 \pm 0.1 $ \\  
$i$          & [deg]                & $ 48.93 \pm 0.97 $    & $  48.07 \pm 2.06 $ & $ 50 $ (fixed)	 \\  
$\Omega$     & [deg]                & $ 0 $ (fixed)         & $  -2.32 \pm 0.94 $ & $  0 $ (fixed) \\  \hline
\noalign{\smallskip}
$a_1 \sin i$ & [$10^{-3}$ AU]       & $ 1.19  $           & $ 0.23  $ 	  & $ 1.2 \times 10^{-3} $ \\
$f (M)$      & [$10^{-9}$ M$_\odot$]& $60.41  $           & $ 1.79  $ 	  & $ 5.8 \times 10^{-5} $ \\
$M \sin i$ & [M$_\oplus$]           & $ - $               & $ - $ 	  & $ 6.3 $ \\
$M$        & [M$_\mathsf{Jup}$]     & $ 2.64  $           & $ 0.83  $ 	  & $ - $ \\
$a$          & [AU]                 & $ 0.211 $           & $ 0.132 $ 	  & $ 0.021 $ \\ \hline
\noalign{\smallskip}
$N_\mathrm{meas}$        &          & \multicolumn{3}{c}{207}   \\
Span         & [day]                & \multicolumn{3}{c}{4103}   \\
$\sqrt{\chi^2}$     &               & \multicolumn{3}{c}{1.37}   \\ 
%$rms_{\rm (averaged)}$ & [m/s]      & \multicolumn{3}{c}{2.29}   \\
$rms_{(Keck)}$         & [m/s]      & \multicolumn{3}{c}{4.25}   \\ %4.2541
$rms_{(HARPS)}$        & [m/s]      & \multicolumn{3}{c}{1.80}   \\ \hline %1.7982
\noalign{\smallskip}
\end{tabular}

Errors are given by the standard deviation $ \sigma $, and $ \lambda $ is the
mean longitude of the date. 
\end{center}
\end{table*}

\begin{figure}
   \centering
    \includegraphics*[width=8.5cm]{\figpath 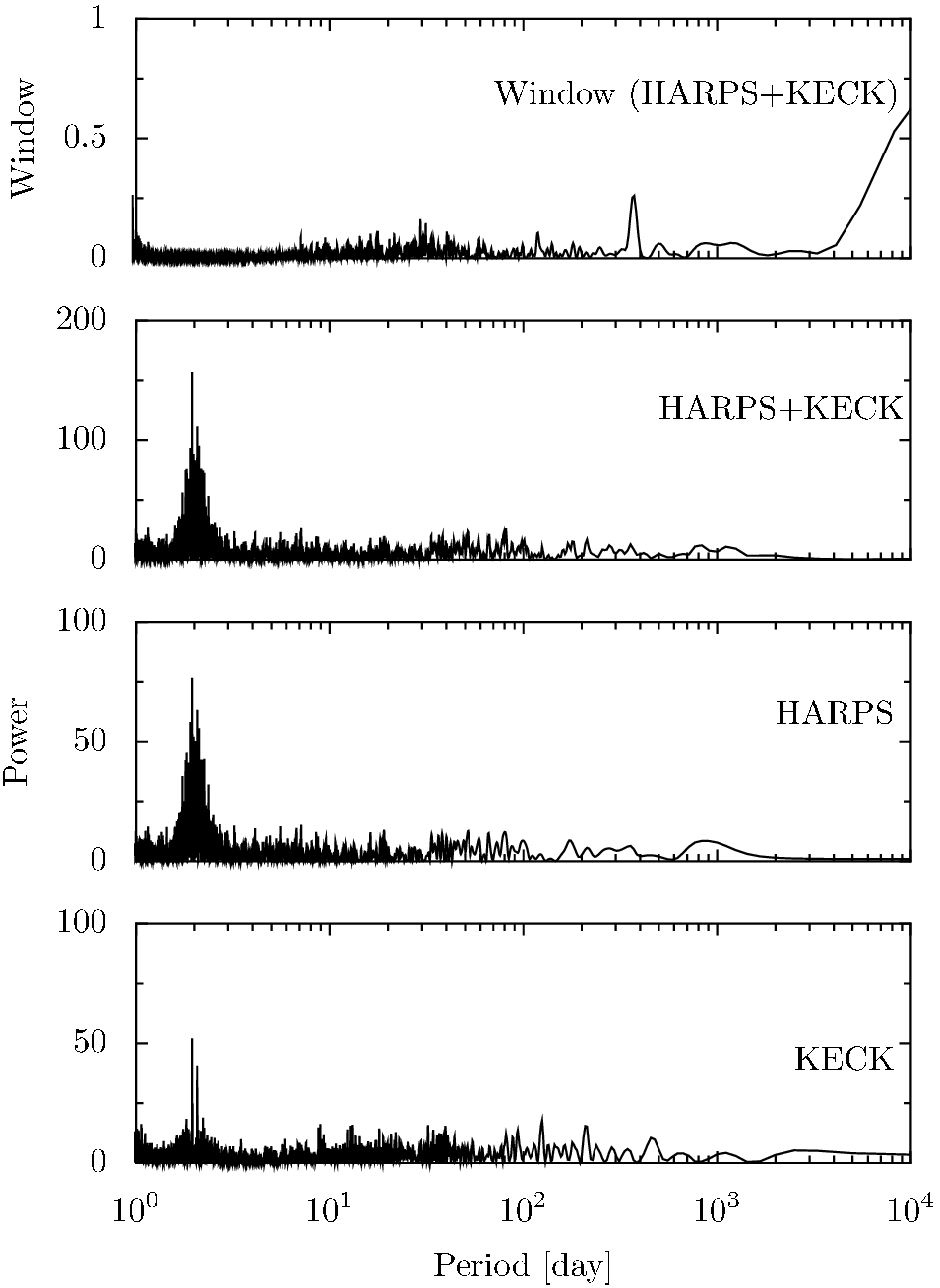}
  \caption{Frequency analysis and window function for the
  two-planet Keck and HARPS residual radial velocities of GJ\,876 (Fig.\ref{F1}).  
  An important peak is detected at $P = 1.9379$\,day, which can be interpreted as
  a third planetary companion in the system. 
  Looking at the window function, we can see that this peak is not an artifact. 
    \label{F2} }
\end{figure}

The residuals around the best-fit two-planet solution are small, but still
larger than the internal errors (Fig.\,\ref{F1}).
We may then search for other companions with different orbital periods.
Performing a frequency analysis of the velocity residuals (Fig.\ref{F1}), we find
an important peak signature at $ P = 1.9379 $~day (Fig.\ref{F2}).
This peak was already present in the Keck data analysis performed by
\citet{Rivera_etal_2005}, but it is here reinforced by the HARPS data.
In Fig.~\ref{F2}, we also plot the window function and conclude that the above
mention peak cannot be an aliasing of the observational data.

To test the planetary nature of the signal, we performed a Keplerian fit
to the residuals of the two planets.
We found an elliptical orbit with $P_d = 1.9378 $~day, $e_d= 0.14$, and
$M_d \sin i_d = 5.4 $~M$_\oplus$. 
The adjustment gives $ \sqrt{\chi^2} = 1.52 $ and
$rms$\,=\,$2.63\,\mathrm{m/s}$, which represents a substantial improvement with
respect to the system with only two companions (Fig.~\ref{F1}).

\begin{figure}
    \includegraphics*[height=8.8cm,angle=270]{\figpath 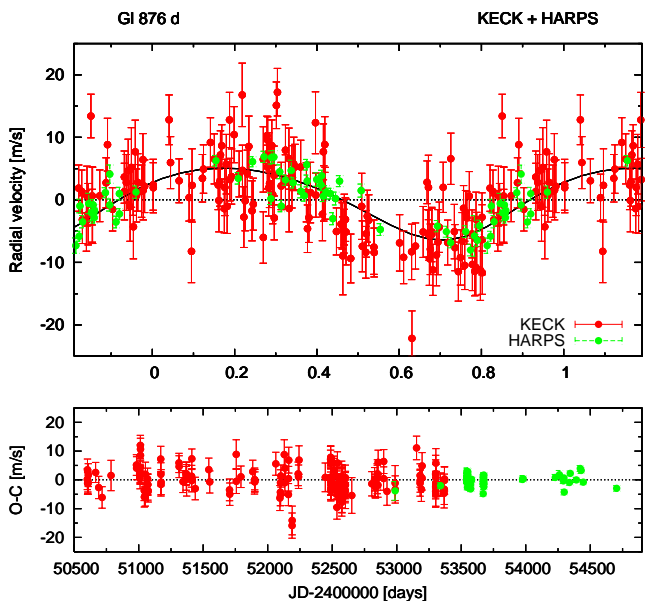} 
  \caption{Phase-folded residual radial velocities for 
    GJ\,876 when the contributions from the planets
    $b$ and $c$ are subtracted. Data are superimposed on a Keplerian orbit
    of $P = 1.9378$~day and $ e = 0.14 $ (the complete set of orbital parameters are
    those from Table\,\ref{T2}). The respective residuals as a function of
    Julian date are displayed in the lower panel.
    \label{F3}}   
\end{figure}

\subsection{Complete orbital solution}

Starting with the orbital parameters derived in the above sections, 
we can now consider the best-fit orbital solution for GJ\,876.
Using the iterative Levenberg-Marquardt method, we then fit the 
complete set of radial velocities with a 4-body Newtonian model.
In this model, 20 parameters out of 23 possible are free to vary.
The three fixed parameters are $\Omega_b = 0^\circ $ 
(by definition of the reference plane), but also $\Omega_d = 0^\circ $ and $ i_d = 50^\circ $,
because the current precision and time span of the observations are not
large enough to
constrain these two orbital parameters. However, the quality of the fit does not
change if we choose other values for these parameters.

The orbital parameters corresponding to the best-fit solution are listed
in Table\,\ref{T2}.
In particular, we find $i_b = 48.9^\circ \pm 1.0^\circ $ and $i_c = 48.1^\circ \pm
2.1^\circ$, which infer, respectively, $M_b = 2.64 \pm 0.04 \,M_{\rm
Jup}$ and $ M_c = 0.83 \pm 0.03 \,M_{\rm Jup}$ for the true masses of the planets.
This fit yields an adjustment of $\sqrt{\chi^2}=1.37$ and $rms=2.29\,\mathrm{m
s}^{-1}$, which represents a significant improvement with respect to all previous
solutions.
We note that the predominant uncertainties are related to the star's mass
(Table\,\ref{T1}), but these are not folded into the quoted error bars.

The best-fit orbit for planet $d$ is also eccentric, in contrast to previous
determinations in which its value was constrained to be zero
\citep{Rivera_etal_2005,Bean_Seifahrt_2009}. 
We can also fix this parameter and our revised solution corresponds to
an adjustment with
$ \sqrt{\chi^2} = 1.40 $ and $rms$\,=\,$2.34\,\mathrm{m/s}$.
These values are compatible with the best-fit solution in Table\,\ref{T2}, so we
cannot rule out the possibility that future observations will decrease the
eccentricity of planet $d$ to a value close to zero.

\subsection{Inclination of planet $d$}

With the presently available data, we were able to obtain with great accuracy the
inclinations of planets $b$ and $c$.
However, this was not the case for planet $d$, whose inclination was held fixed at
$50^\circ$ in the best-fit solution (Table\,\ref{T2}).
We are still unable to determine the inclination of the innermost planet, because the
gravitational interactions between this planet and the other two are not as
strong as the mutual interactions between the two outermost planets.

\begin{figure}
    \includegraphics*[width=8.8cm]{\figpath 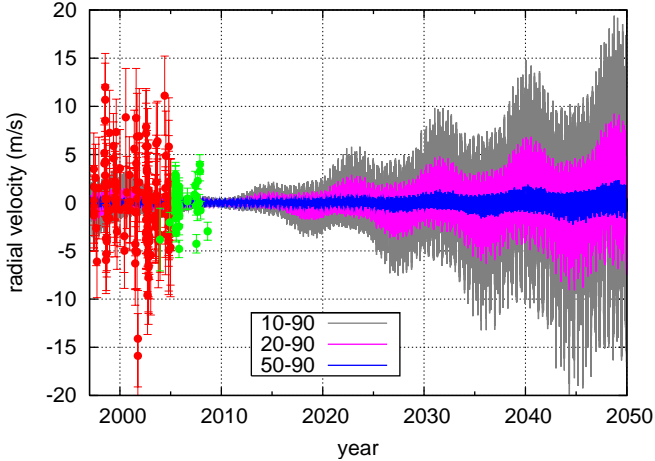} 
  \caption{Radial velocity differences between orbital solutions with $i_d =
  10^\circ $, $20^\circ$, and $50^\circ$ and the orbital solution with $i_d =
  90^\circ$. We also plot the velocity residuals of the best-fit solution from
  Table\,\ref{T2}. We observe that, with the current HARPS precision, the
  inclination will be constrained around 2025 if $i_d$ is close to $10^\circ$, but
  if this planet is also coplanar with the other two ($i_d \sim 50^\circ $), its
  precise value cannot be obtained before 2050. \label{F4}}   
\end{figure}

We can nevertheless estimate 
the  time span of GJ\,876  radial velocity measures that is necessary 
before we will be able to determine the inclination $i_d$ with good accuracy.
For that purpose, we fit the observational data for different fixed values of
the inclination of planet $d$ ($i_d = 10^\circ$, $20^\circ$, $50^\circ$,
and $90^\circ$). 
In Fig.\,\ref{F4}, we plot the evolution of the radial velocity differences
between each solution and the solution at $90^\circ$.
We also plot the velocity residuals of the best-fit solution from
Table\,\ref{T2}.
We observe that, with the current HARPS precision, the inclination will be
constrained around 2025 if $i_d$ is close to $10^\circ$, but if this planet is also
coplanar with the other two ($i_d \sim 50^\circ $), its precise value cannot be
obtained before 2050 (Fig.\,\ref{F4}).

\subsection{Other instruments}

\begin{table*}
\caption{Orbital parameters for the planets orbiting {GJ}\,876, obtained
with a 4-body Newtonian fit to all available radial velocity observational data
(Lick, Keck, CORALIE, ELODIE, and HARPS).  \label{T3} }
\begin{center}
\begin{tabular}{l l c c c} \hline \hline
\noalign{\smallskip}
{\bf Param.}  & {\bf [unit]} & {\bf b} & {\bf c} & {\bf d} \\ \hline 
\noalign{\smallskip}
Date         & [JD]             & \multicolumn{3}{c}{2\,455\,000.00 (fixed)}  \\ 
$V_{Lick}$          & [km/s]    & \multicolumn{3}{c}{$-0.0304 \pm 0.0058 $}  \\ 
$V_{Keck}$          & [km/s]    & \multicolumn{3}{c}{$\phantom{-}0.0131 \pm 0.0058 $}  \\ 
$V_{CORALIE}$       & [km/s]    & \multicolumn{3}{c}{$-1.8990 \pm 0.0061 $}  \\ 
$V_{ELODIE}$        & [km/s]    & \multicolumn{3}{c}{$-1.8624 \pm 0.0064 $}  \\ 
$V_{HARPS}$         & [km/s]    & \multicolumn{3}{c}{$-1.3389 \pm 0.0058 $}  \\ \hline  
\noalign{\smallskip}
$P$          & [day]                & $ 61.065 \pm 0.012 $ & $ 30.259 \pm 0.010 $ & $ 1.93785 \pm 0.00002  $ \\ 
$\lambda$    & [deg]                & $ 35.56  \pm 0.15  $ & $ 159.05 \pm 0.76  $ & $   29.37 \pm 3.21  $ \\ 
$e$          &                      & $  0.031 \pm 0.001 $ & $  0.265 \pm 0.002 $ & $   0.124 \pm 0.032 $ \\ 
$\omega$     & [deg]                & $ 274.69 \pm 2.57  $ & $ 274.64 \pm 1.19  $ & $  168.45 \pm 17.21 $ \\ 
$K$          & [m/s]                & $ 212.09 \pm 0.33  $ & $  85.91 \pm 0.40  $ & $    6.60 \pm 0.26  $ \\  
%$T$          & [JD-2400000]         & $ 54979.5 \pm 0.4 $ & $ 54979.3 \pm 0.1 $ & $ 54998.8 \pm 0.1 $ \\  
$i$          & [deg]                & $ 48.98 \pm 0.94 $    & $  45.67 \pm 1.81 $ & $ 50 $ (fixed)	 \\  
$\Omega$     & [deg]                & $ 0 $ (fixed)         & $  -1.67 \pm 0.84 $ & $  0 $ (fixed) \\  \hline
\noalign{\smallskip}
$a_1 \sin i$ & [$10^{-3}$ AU]       & $ 1.19  $           & $ 0.23  $ 	  & $ 1.2 \times 10^{-3} $ \\
$f (M)$      & [$10^{-9}$ M$_\odot$]& $60.27  $           & $ 1.78  $ 	  & $ 5.6 \times 10^{-5} $ \\
$M \sin i$ & [M$_\oplus$]           & $ - $               & $ - $ 	  & $ 6.2 $ \\
$M$        & [M$_\mathsf{Jup}$]     & $ 2.64  $           & $ 0.86  $ 	  & $ - $ \\
$a$          & [AU]                 & $ 0.211 $           & $ 0.132 $ 	  & $ 0.021 $ \\ \hline
\noalign{\smallskip}
$N_\mathrm{meas}$        &          & \multicolumn{3}{c}{309}   \\
Span         & [day]                & \multicolumn{3}{c}{5025}   \\ %5025.086788
$\sqrt{\chi^2}$     &               & \multicolumn{3}{c}{1.46}   \\ 
$rms$        & [m/s]                & \multicolumn{3}{c}{3.01}   \\ \hline
\noalign{\smallskip}
\end{tabular}

Errors are given by the standard deviation $ \sigma $, and $ \lambda $ is the
mean longitude of the date. 
\end{center}
\end{table*}

Besides the Keck and HARPS programs, the star GJ\,876 was also followed by 
many other instruments.
The oldest observational data were acquired using the Hamilton echelle spectrometer mounted on
the 3-m Shane %{\bf (?? correct?)} 
telescope at the Lick Observatory.
The star was followed from November 1994 until December 2000 and 16 radial
velocity measurements were acquired with an average precision of 25~m/s
\citep{Marcy_etal_2001}. 
During 1998, between July and September a quick series of 40 radial velocity
observations were performed using the CORALIE echelle spectrograph
mounted on the 1.2-m Swiss telescope at La Silla, with an average precision of
30~m/s \citep{Delfosse_etal_1998}.
Finally, from October 1995 to October 2003, the star was also observed at the
Haute-Provence Observatory (OHP, France) using the ELODIE high-precision fiber-fed
echelle spectrograph mounted at the Cassegrain focus of the 1.93-m telescope. 
Forty-six radial velocity measurements were taken with an average precision of 18~m/s
(data also provided with this paper via CDS).

We did not consider these data sets in the previous analysis because their
inclusion would have been more distracting than profitable. 
Indeed, the internal errors of these three instruments are much higher than those
from Keck and HARPS series, and they are unable to help in constraining the
inclinations.
Moreover, all the radial velocities are relative in nature, and therefore each
data set included requires the addition of a free offset parameter in the orbit
fitting procedure, $V_{instrument}$.

Nevertheless, to be sure that there is no gain in including these additional 102
measurements, we performed an adjustment to the data using the five instruments
simultaneously. 
The orbital parameters corresponding to the   best-fit solution are listed
in Table\,\ref{T3}.
As expected, this fit yields an adjustment of $\sqrt{\chi^2}=1.46$ and
$rms=3.01\,\mathrm{m s}^{-1}$, which does not represent an improvement
with respect to the fit listed in Table\,\ref{T2}.
The inclination of planet $c$ decreases by $2.5^\circ$, but this difference
is within the $3 \sigma$   uncertainty of the best-fit values.
Therefore, the orbital parameters determined only with data from Keck and HARPS
will be adopted  (Table\,\ref{T2}).

%________________________________________________________________

\section{Dynamical analysis}
\label{dynevol}

We now analyze the dynamics and stability of the
planetary system given in Table\,\ref{T2}.
Because of the two outermost planets' proximity and high values of their masses,
both planets are affected by strong planetary perturbations from each other.
As a consequence, unless a resonant mechanism is present to avoid close encounters, the
system cannot be stable. 

\subsection{Secular coupling}

As usual in planetary systems, there is  a strong  coupling 
within the secular system \citep[see][]{Laskar_1990}. Therefore, both planets $b$ and $c$ precess with 
the same precession frequency $g_2$, which is retrograde with a period of 8.74~years. 
The two periastron are thus locked and $\Delta \varpi = \varpi_c-\varpi_b$
oscillate around $0^\circ$ with an amplitude of about  $25^\circ$ (Fig.\,\ref{F5}).
This behavior, already mentioned in  earlier studies 
\citep{Laughlin_Chambers_2001,Lee_Peale_2002,Beauge_etal_2003}, 
is not  a dynamical 
resonance, but merely the result of the linear secular coupling. 
To present the solution in a clearer way, it is then useful to 
make a linear change of variables into eccentricity and inclination 
proper modes \citep[see][]{Laskar_1990}. 
In the present case, the linear transformation is numerically 
obtained by a frequency analysis of the solutions.
For the dynamical analysis, to understand the evolution of the inclinations in this system, 
 we have expressed all the coordinates 
in the reference frame of the invariant plane,
orthogonal to the total angular 
momentum of the system.  The longitude of node $\Omega$ and inclination $I$ 
of this  invariant plane are
\begin{equation}
\Omega =-0.445256^\circ;\quad I = 48.767266^\circ \ .
\end{equation}
Using the classical complex notation,   
\begin{equation}
z_k = e_k \ex^{i \varpi_k} ; \quad \zeta_k = \sin(i_k/2) \ex^{i \Omega_k} ;
\end{equation}
for $k=b,c,d$, we have for the linear Laplace-Lagrange solution
\begin{equation}
\left(\begin{array}{c} z_d\\z_c\\z_b \end{array}\right)= 
\left(\begin{array}{rrr} 
 0.139690&      0.000843&   -0.001265\\
-0.000069&      0.265688&   -0.006501\\
 0.000063&      0.037609&    0.006970\\
\end{array}\right)
\left(\begin{array}{c} u_1\\u_2\\u_3 \end{array}\right) \ ,
\label{eq.lape}
\end{equation}
where the proper modes $u_k$ are obtained from the $z_k$ 
by inverting the above linear relation. To good 
approximation, we then have 
$u_k \approx  \ex^{i (g_k t +\phi_k)}$, where $g_k$ and $\phi_k$ are given in Table\,\ref{tab.freq}.
For the inclinations, due to the conservation of angular momentum, 
there are only two proper modes, $v_1, v_2$, which we can define with the 
invertible linear relation 

\begin{equation}
\left(\begin{array}{c} \zeta_d\\ \zeta_c\end{array}\right)= 
\left(\begin{array}{rr} 
 0.009449& +0.002723\\
-0.000004& -0.013783\\
\end{array}\right)
\left(\begin{array}{c} v_1\\v_2\end{array}\right) \ ,
\label{eq.lapi}
\end{equation}
and the additional approximate relation 
\begin{equation}
\zeta_b  \approx  -0.000028\, v_1 +0.003301\, v_2 \ .
\label{eq.lapib}
\end{equation}
The proper modes in inclination are then given to a good approximation as 
$v_k \approx \ex^{i (s_k t +\psi_k)}$, where $s_k$ and $\psi_k$ are given in Table\,\ref{tab.freq}. 

\begin{figure}
    \includegraphics*[width=8.8cm]{\figpath 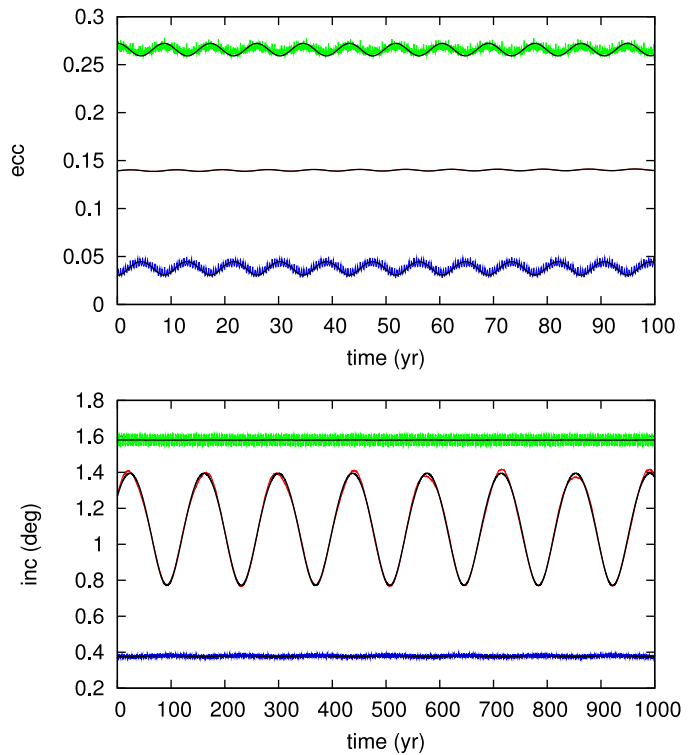} \\    
  \caption{Evolution of the {GJ}\,876 eccentricities (top) and inclinations
  (bottom) with time, starting with the orbital solution from Table\,\ref{T2}.
  The color lines are the complete solutions for the various planets (b: blue, c: green,
  d: red), while the black curves are the associated values obtained with 
  the linear, secular model. 
   \label{F9}}   
\end{figure}

With Eq.\,\ref{eq.lape}, it is then easy to understand the meaning of the observed 
libration between the periastrons $\varpi_b$ and $\varpi_c$. Indeed, for both planets 
$b$ and $c$, the dominant term is the $u_2$ term with frequency $g_2$. They 
thus both precess with an average value of $g_2$. 
In the same way, both nodes $\Omega_b$ and $\Omega_c$ precess with the same 
frequency $s_2$.
It should also be noted that Eqs. \ref{eq.lape}, \ref{eq.lapi}, and
\ref{eq.lapib}
provide good approximations of the long-term evolution of the 
eccentricities and inclinations. Indeed, in Fig.\,\ref{F9} we plot the eccentricity 
and the inclination with respect to the invariant plane of planets $b$, $c$, $d$, 
with initial conditions from Table \ref{T2}. At the same time, we plot with solid black
lines the evolution of the same elements given by the above secular,
linear approximation.  

The eccentricity and inclination variations are very limited and described well by 
the secular approximation. 
The eccentricity of planets $b$ and $c$ are within the ranges $ 0.028 < e_b < 0.050 $ and $
0.258 < e_c < 0.279 $, respectively. These variations are  
driven mostly by  the rapid secular frequency $g_3 - g_2$, of period $ 2 \pi / (g_3 - g_2)
\approx 8.62 $~yr (Table\,\ref{T4}). The eccentricity of planet $d$ 
is nearly constant with $ 0.136 < e_d < 0.142 $.

The inclinations of $b$ and $c$ with respect to the invariant plane are very small with
$ 0.36^\circ <i_b < 0.39^\circ $ and $ 1.54^\circ < i_c < 1.61^\circ $, respectively.
This small variation is mostly caused by the non-linear secular term $2
(s_2-g_2) $  of period 4.791~yr. 
Although the inclination of $d$ is not well constrained, using the initial conditions 
of Table \ref{T2}, one finds small variations in its inclination 
$ 0.76^\circ < i_c < 1.42^\circ $, which are driven by the secular frequency 
$s_1-s_2$ of period 138.3~yr (Table \ref{T4}).

With the present 11~years of observations covered by Keck and HARPS,
the most important features that allow us to constrain the  parameters
of the system are those related  to the rapid  secular frequencies 
$g_2$ and $s_2$, of periods 8.7~yr and 99~yr, which are the precession 
frequencies of the periastrons and nodes of planets $b$ and $c$.

\subsection{The 2:1 mean motion resonance}

\begin{table}
 \caption{Fundamental frequencies and phases for the orbital solution in
 Table\,\ref{T2}. 
 \label{T4}} 
 \begin{center}
 \begin{tabular}{crr r r}
 \hline\hline
      & Frequency   & Period &\multicolumn{2}{r}{Phase}\\
      & deg/yr      & yr & \multicolumn{2}{r}{deg}\\
 \hline
      $n_b$ &  2154.322532 &   0.167106 & $\quad \quad \lambda_{b0}$ & 36.0925 \\
      $n_c$ &  4349.841605 &   0.082762 & $\lambda_{c0}$ &158.1879\\
      $n_d$ & 67852.886097 &   0.005306 & $\lambda_{d0}$ & 29.6886\\
      $g_1$ &     1.003492 & 358.747259 & $\phi_1$ &170.0005 \\
      $g_2$ &   -41.196550 &   8.738596 & $\phi_2$ &-86.0030 \\
      $g_3$ &     0.555047 & 648.593233 & $\phi_3$ & 89.8880 \\
      $s_1$ &    -1.021110 & 352.557387 & $\psi_1$ &  3.9446\\
      $s_2$ &    -3.624680 &  99.319103 & $\psi_2$ & 64.1066\\
 $l_\theta$ &   245.918001 &   1.463903 & $\theta_0$ & -48.6812\\ 
 \hline
\end{tabular}
\end{center}
$n_b$, $n_c$ and $n_d$ are the mean motions, $g_1$, $g_2$ and $g_3$ the main
 secular frequencies of the periastrons, $s_1$ and $s_2$ the main secular
 frequencies of the nodes, and $l_\theta$ the libration frequency of the resonant angles. 
 %$\theta_b $ and $\theta_c $. Indeed, we have $ 2 n_b - n_c - g_2 = 0 $.
 \label{tab.freq}
\end{table}

The ratio of the orbital periods of the two outermost planets determined 
by the fitting process (Table\,\ref{T2}) is $ P_b
/ P_c = 2.018 $, suggesting that the system may be trapped in a 2:1
mean motion resonance. 
This resonant motion has already been reported in previous works
\citep{Laughlin_Chambers_2001,Rivera_Lissauer_2001,Lee_Peale_2002,Rivera_etal_2005}.
To test the accuracy of this scenario, we performed a frequency analysis of the 
orbital solution listed in Table\,\ref{T2} computed over 100~kyr.
The orbits of the three planets are integrated with the 
symplectic integrator SABA4 of \citet{Laskar_Robutel_2001}, using a step size
of $2 \times 10^{-4}$~years.
We conclude that in the nominal solution of Table\,\ref{T2},
planets $b$ and $c$ in the {GJ}\,876 system indeed show a
2:1 mean motion resonance, with resonant arguments:
$\theta_b = 2 \lambda_b - \lambda_c - \varpi_b $
and
$\theta_c = 2 \lambda_b - \lambda_c - \varpi_c$.

\begin{figure}
    \includegraphics*[width=8.8cm]{\figpath 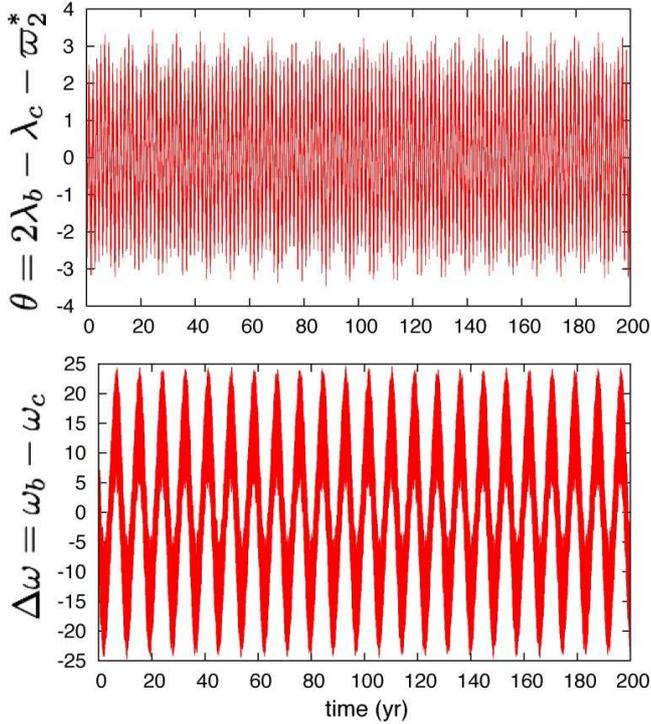} 
    % use file reson.png or pdf for this figure
  \caption{Variation in the resonant argument $\theta= 2 \lambda_b - 
  \lambda_c - \varpi^*_2 $  with time (top). The difference 
  $\Delta \omega = \varpi_b-\varpi_c$  also oscillates around $0^\circ$ with a 
  $25^\circ$ amplitude.}
  \label{F5}   
\end{figure}

If we analyze these arguments, it is indeed difficult to 
disentangle the proper libration from the secular oscillation of the periastrons
angles $\varpi_b, \varpi_c$. It is thus much clearer to switch to proper modes $u_k,v_k$
by means of the linear transformation (\ref{eq.lape}, \ref{eq.lapi}).
Indeed, if $u_k=\ex^{i \varpi^*_k}$, 
the argument 
\begin{equation}
\theta = 2 \lambda_b - \lambda_c + \varpi^*_2  \label{rarg}
\end{equation}
is in  libration around $0^\circ$ with a very small total amplitude 
of $3.5^\circ$ (Fig.\,\ref{F5})  with only $1.8^\circ$ amplitude 
related to the proper libration
argument $\theta$
with   period   $ 2 \pi / l_\theta = 1.46 $~yr. The remaining terms of 
Table \ref{Tj5}  are forced terms related to planetary interactions.
The fundamental frequencies associated with this libration argument  
(Table\,\ref{T4}) indeed verify the resonance relation 
$2 n_b- n_c - g_2 = 0 $
up to the precision of the determination of the frequencies ($\approx 10^{-5}$).

%%%%%%%%%%%

\begin{table}
\caption{Quasi-periodic decomposition of the resonant angle
$\theta = 2 \lambda_b - \lambda_c - \varpi_2^*$
for an integration  over 100~kyr of the orbital solution in
  Table\,\ref{T2}. \label{T5}} 
  \begin{center}
    \begin{tabular}{rrrrrrrr}
\hline\hline
\multicolumn{5}{c}{\textbf{Combination}} & $\nu_i$ & $A_i$  & $\phi_i$  \\
      $n_b$ & $n_c$ & $g_2$&$g_3$&  $l_\theta$ & (deg/yr) & (deg) & (deg) \\
\hline
   0 &  0 &   0  &  0 &    1 &      245.9180 &         1.810 &      -48.681 \\ 
   1 &  1 &  -2  &  0 &    0 &     6586.5572 &         0.567 &       96.286 \\ 
   0 &  1 &  -1  &  0 &    0 &     2195.5191 &         0.569 &       32.095 \\ 
   0 &  0 &  -1  &  1 &    0 &       41.7516 &         0.411 &      -94.109 \\ 
   1 &  0 &  -1  &  0 &    0 &     4391.0382 &         0.255 &      -25.809 \\ 
   2 &  1 &  -3  &  0 &    0 &    10977.5954 &         0.156 &      -19.523 \\ 
   0 &  0 &   1  & -1 &    1 &      204.1664 &         0.120 &      -44.572 \\ 
  \hline
\end{tabular}
\end{center}
  We have $\theta = \sum_{i=1}^N A_i \cos(\nu_i\, t + \phi_i)$. 
  We only provide the first 7 largest terms that are identified as
  integer combinations 
  of the fundamental frequencies given in
  Table\,\ref{T4}. 
  \label{Tj5}
\end{table}

\subsection{Coplanar motion}

The orbits of the planets in the Solar System are nearly coplanar, that is, their
orbital planes remain within a few degrees of inclination from an inertial plane
perpendicular to the total angular momentum of the system. The highest
inclination is obtained for Mercury ($I \sim 7^\circ$), which is also the smallest of the
eight planets.

A general belief that planetary systems would tend to resemble the
Solar System persists, but there is no particular reason for all
planetary systems to be coplanar such as our own.
Star formation theories require an accretion disk in which planets
form, but close encounters during the formation process can increase the
eccentricities and inclinations \citep[e.g.][]{Lee_etal_2007}.
That planets are found to have significant eccentricity values is
also consistent with their having high inclinations \citep[e.g.][]{Libert_Tsiganis_2009}.
In particular, studies of Kuiper belt objects indicate that there is an important
inclination excitation when the bodies sweep secular resonances
\citep[e.g.][]{Nagasawa_etal_2000, Li_etal_2008}, a mechanism that also appears
to be applicable to planets \citep[e.g.][]{Thommes_Lissauer_2003}.

The question of whether extra-solar planetary systems are also nearly coplanar
or not is thus important.
Although about 45 extra-solar multi-planet systems have been reported,
their true inclinations have so far been determined in only two cases.
One is the planetary system around the pulsar PSR\,B1257\,+\,12, discovered by
precise timing measurements of pulses \citep{Wolszczan_Frail_1992}, for which
the orbits seem to be almost coplanar \citep{Konacki_Wolszczan_2003}.
The other case is GJ\,876, the only planetary system around a main-sequence
star for which one can access the inclinations.
This result  is  possible  because of the large amount of
data already available and because the two giant planets show strong 
gravitational  interactions.

\begin{figure}
  \centering
    \includegraphics*[width=9cm]{\figpath 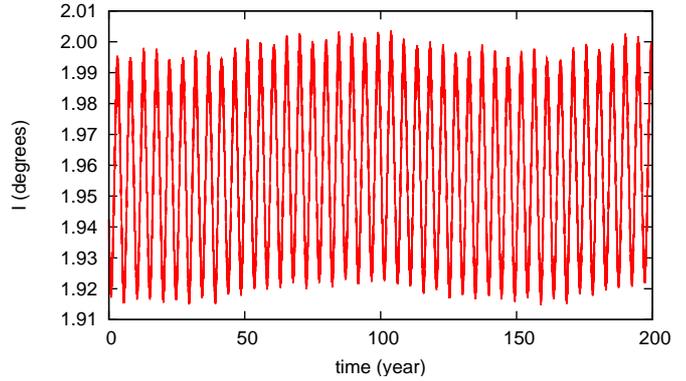} 
  \caption{Evolution of the mutual inclination of the planets $b$ and $c$. Since
  the maximal inclination between the two planets is $2.0^\circ$, we
  conclude that the orbits are almost coplanar, as in our Solar System.
  \label{F6}}   
\end{figure}

The analytical expressions and the plots versus time of the orbital elements 
in reference frame of the invariant 
plane  show that the  inclinations of planets $b$ and $c$
are both very small (Eqs.\,\ref{eq.lapi}, \ref{eq.lapib}, Fig.\,\ref{F9}). 
To test the coplanarity of this system, we plot in Fig.~\ref{F6}
the evolution of the mutual inclination with time.
We verify that $ I = 1.958^\circ \pm 0.045^\circ$, and hence conclude that
the orbits of planets $b$ and $c$ are indeed nearly coplanar.

In our best-fit solution (Table\,\ref{T2}), we also assumed that the orbit of
planet $d$ is nearly coplanar with the other two planets, but there is no clear
physical reason that justifies our choice.
Indeed, its orbit should be more perturbed by the oblateness of the star, rather
than the other planets \citep{Goldreich_1965,Correia_2009}. 
It is therefore more likely
that planet $d$ orbits in the equatorial plane of the star. 
This plane can be identical to the orbital plane
of the outer planets or not, depending on the evolutionary process.
According to Fig.\,\ref{F2}, with the current HARPS precision of $\sim$1~m/s,
we will have to wait until 
about 2050 to confirm an inclination close to $50^\circ$.

Adopting $\Omega_d = 0^\circ$, and initial values of $i_d$ ranging from $10^\circ$ to
$90^\circ$, we found best-fit solutions with very similar values of reduced $\sqrt{\chi^2}$ as
the solution in Table\,~\ref{T2}. For inclinations lower than $10^\circ$, the 
best-fit solution becomes worse ($\sqrt{\chi^2} = 1.58$ for $i_d=5^\circ$) and
infers low eccentricity values for planet $d$. In addition, planets $b$ and $c$
are no longer nearly coplanar. 
We therefore conclude that a lower limit to the inclination of
planet $d$ can be set to be around $10^\circ$. 
On the other hand, a lack of transit detections only allows inclinations close
to $90^\circ$ if the planet is very dense \citep{Rivera_etal_2005}.

\subsection{Stability analysis}

\begin{figure}
  \centering
   \includegraphics[width=8.8cm]{\figpath 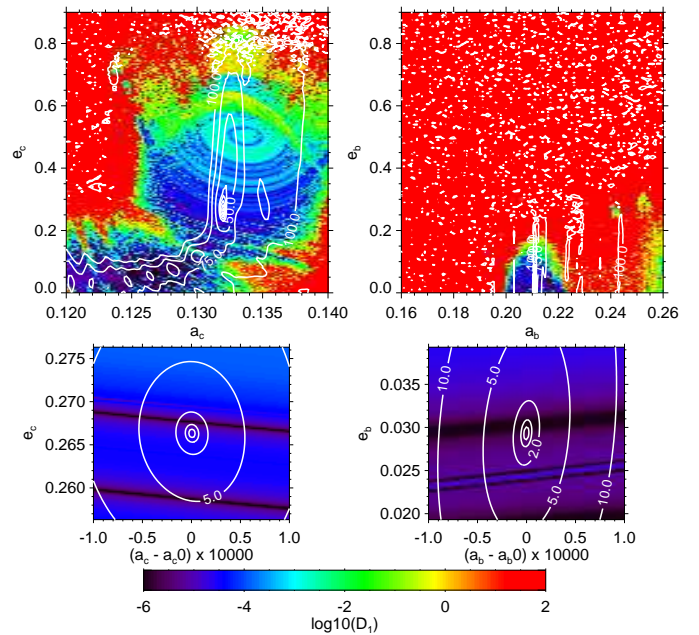}
  \caption{
  Stability analysis of the nominal fit (Table\,\ref{T2}) 
  of the {GJ}\,876 planetary system. For a fixed initial condition 
  of planet $b$ (left) and planet $c$ (right), the phase space of the 
  system is explored by varying the semi-major axis $a_k$ and eccentricity 
  $e_k$ of the other planet, respectively. The step size is $0.0002$~AU in semi-major axis 
  and $0.005$ in eccentricity in the top panels and $2 \times 10^{-6}$~AU and
  $2 \times 10^{-4}$ in bottom panels.
  For each initial condition, the system is integrated over 800~yr
  with an averaged model for planet $d$ \citep{Farago_etal_2009}
  and a stability criterion is derived  with the frequency analysis of the mean longitude
  \citep{Laskar_1990,Laskar_1993PD}.
  As in \citet{Correia_etal_2005,Correia_etal_2009}, the chaotic diffusion 
  is measured by the variation in the frequencies. The ``red'' zone corresponds to highly unstable 
  orbits, while  the ``dark blue'' region can be assumed to be stable on a
  billion-years timescale.
  The contour curves indicate the value of $\chi^2$ %$\sqrt{\chi^2}$ 
  obtained for each choice of parameters.
  It is remarkable that in the present fit, there is an exact 
  correspondence between the  zone of minimal $\chi^2$ %$\sqrt{\chi^2}$, 
  and the 2:1 stable resonant zone, in ``dark blue''.
  \label{F7}}   
\end{figure}

To analyze the stability of the nominal solution (Table\,\ref{T2}) and
confirm the presence of the 2:1 resonance, 
we performed a global frequency analysis \citep{Laskar_1993PD} 
in the vicinity of this solution (Fig.\,\ref{F4}), in the same way as
achieved for the {HD}\,45364 system by \citet{Correia_etal_2009}.

For each planet, the system is integrated on a regular 2D mesh of initial conditions, 
with varying semi-major axis and eccentricity, while the other parameters are 
retained at their nominal values (Table\,\ref{T2}). The solution is integrated over 800~yr 
for each initial condition and a stability indicator is computed 
to be the variation in the measured mean motion over the two consecutive 
400~yr intervals of time. For regular motion, there is no significant variation 
in the mean motion along the trajectory, while it can vary significantly 
for chaotic trajectories. The result is reported in color in Fig.\,\ref{F7}, 
where ``red'' represents the strongly chaotic trajectories, and ``dark blue'' 
the extremely stable ones. To decrease the computation time, we averaged the
orbit of planet $d$ over its mean-motion and periastron, following
\cite{Farago_etal_2009}. 

In the two top plots (Figs. \ref{F7}a,b),
the only stable zone that exists in the vicinity of the 
nominal solution corresponds to the stable 2:1 resonant areas. 
As for HD\,45364 \citep{Correia_etal_2009} and HD\,60532
\citep{Laskar_Correia_2009} planetary systems, there is a perfect coincidence
between the stable 2:1 resonant islands, and curves of minimal $\chi^2$
obtained by comparison with the observations.
Since these islands are the only stable zones in the vicinity, this
picture presents a very coherent view of dynamical analysis and radial
velocity measurements, which reinforces the confidence that the present system 
is in a 2:1 resonant state.  

The scale of the two bottom panels shows the remarkable precision of the data
in light of the dynamical environment of the system. 
The darker structures in these plots can be identified as secondary resonances.
For instance, in the bottom left plot of Fig. 7, the two dark horizontal lines
starting at $e_c \simeq 0.269$ and $0.260$ correspond, respectively, to $l_\theta +
6g_2 - 6g_3 + s_1 - s_2$ and $l_\theta + 6g_2 - 2g_3 - 4s_2$. We can also barely
see the secondary resonance $l_\theta + 6g_2 - 6g_3$ just above the former one.
In the bottom right plot, the bluer line starting at $e_b \simeq 0.023$
corresponds to  $l_\theta + 6g_2 - 4g_3 - 2s_2$. A longer integration time
would most probably provide more details about the resonant and secular dynamics.

\subsection{Long-term orbital evolution}

\begin{figure}
    \includegraphics*[width=8.8cm]{\figpath 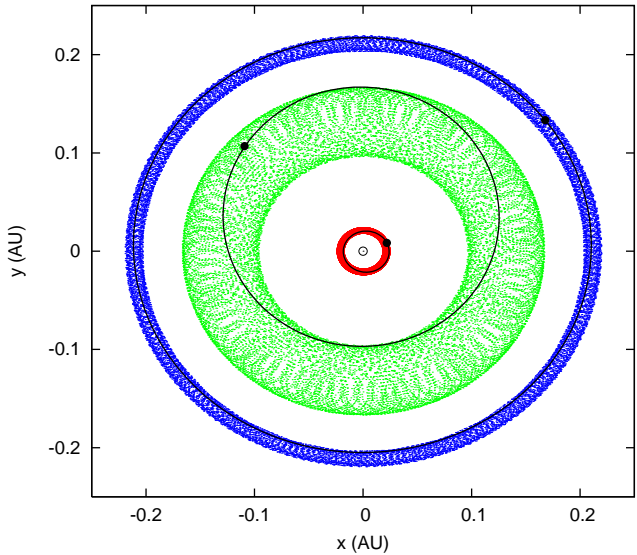} 
  \caption{Long-term evolution of the {GJ}\,876 planetary system over 100~Myr
  starting with the orbital solution from Table\,\ref{T2}. We did not include
  tidal effects in this simulation. 
  The panel shows a face-on view of the system invariant plane. $x$ and $y$ are spatial
  coordinates in a frame centered on the star. Present orbital solutions are traced
  with solid lines and each dot corresponds to the position of the planet every
  5~kyr. The semi-major axes (in AU) are almost constant ($ 0.209 < a_b < 0.214 $;
  $ 0.131 < a_c < 0.133 $ and $ 0.02110 < a_d < 0.02111 $), and the eccentricities
  present slight variations ($ 0.028 < e_b < 0.050 $; $
  0.258 < e_c < 0.279 $ and $ 0.136 < e_d < 0.142 $). 
  %The fundamental period related to the precession of the periastrons is $ 2 \pi / g_2 = - 8.74$~yr.
  \label{F8}}   
\end{figure}

\begin{figure*}
  \centering
    \includegraphics[width=16cm]{\figpath 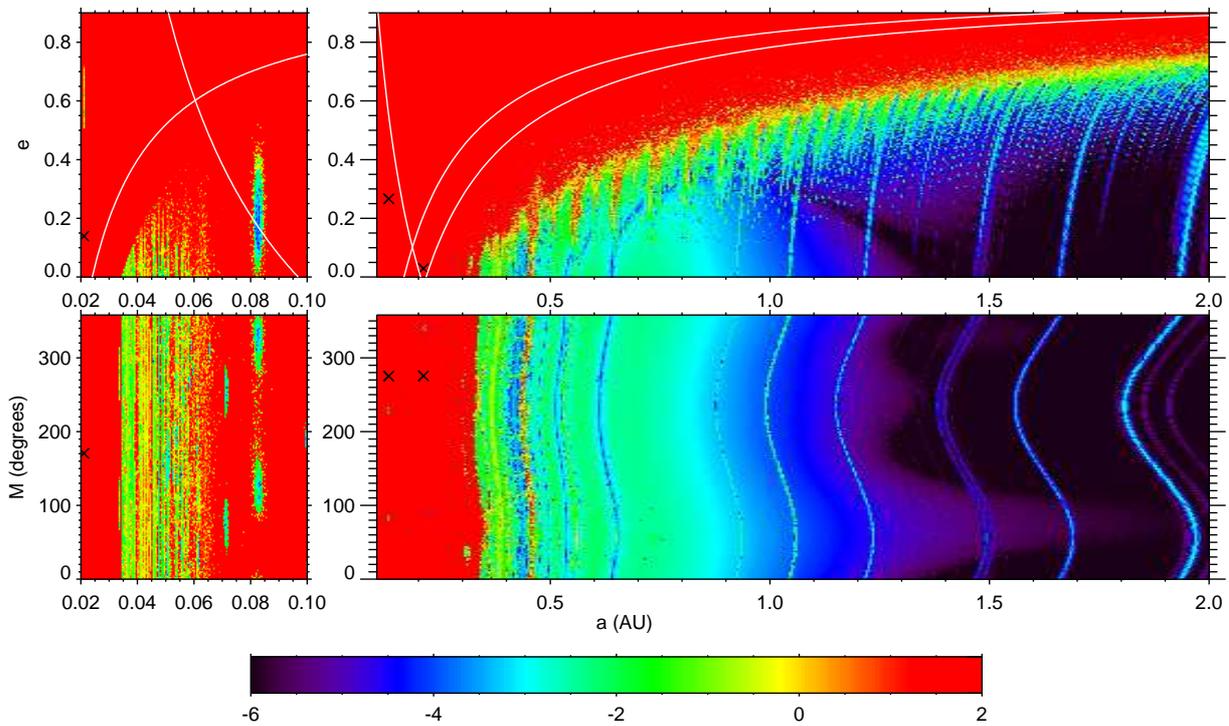} 
  \caption{Possible location of an additional fourth planet in the GJ\,876 system.
  The stability of an Earth-size 
  planet ($K=1$~m/s) is analyzed, for various semi-major axis versus 
  eccentricity (top), or mean anomaly (bottom). All the angles of the putative planets are set to $0^\circ$
  (except the mean anomaly in the bottom panels),
  its inclination to $50^\circ$, and in the bottom panels, its eccentricity to 0.
  The stable zones  where additional planets
  can be found are the dark blue regions. Between the already known planets,
  we can find stability in a small zone corresponding to a 4:1 mean motion
  resonance with planet $b$ (and 2:1 resonance with planet $c$). White lines
  represent the collisions with the already existing planets, given by a cross.
    \label{F10}}   
\end{figure*}

From the previous stability analysis, it is clear that planets $b$ and $c$
listed in Table\,\ref{T2} are trapped in  
 a 2:1 mean motion resonance  and stable over a Gyr timescale.
Nevertheless, we also tested this by performing a numerical integration
of the orbits using the symplectic integrator SABA4.

Because the orbital period of the innermost planet is shorter than 2~day, we
performed two kinds of experiments.
In the first one, we directly integrated the full planetary system 
over 100~Myr  with a step size of $2 \times 10^{-4}$~years.
Although tidal effects may play an important role in this system evolution, we
did not include them.
The result is displayed in Fig.\,\ref{F8}, showing that the orbits indeed 
evolve in a regular way, and remain stable throughout the simulation.
For longer timescales, we needed to average the orbit of the inner planet as
explained in \citet{Farago_etal_2009}.
We then use a longer step size of $2 \times 10^{-3}$~years and integrated the
system over 5~Gyr, which corresponds to the maximal estimated age of
the central star. The subsystem consisting of the giant planets $b$ and $c$
remained stable. 

In spite of the strong gravitational interactions between the two planets locked
in the 2:1 mean motion resonance, both orbital eccentricities and inclinations
exhibit small variations that are mostly driven by the regular linear secular 
terms (Fig.\ref{F9}).
These variations occur far more rapidly than in our Solar System, which enabled
their direct detection using only 11~yr of data taken with Keck and HARPS together.
It is important to notice, however, that
for initial inclinations of planet $d$ very different from $50^\circ$, the
orbital perturbations from the outer planets may become important.
For instance, using an orbital solution with $i_d = 90^\circ $, we derive an
eccentricity and inclination variations of $ 0.10 < e_d < 0.35 $ and $ 10^\circ <
i_d < 90^\circ $.

\subsection{Additional constraints}

The stability analysis summarized in Fig. \ref{F7} shows good agreement
between  the 2:1 resonant islands and the $\chi^2$ contour curves.
We can thus assume that the dynamics of the two known planets is not 
disturbed much by the presence of an additional planet close-by.
The same is true for the innermost planet, which has an orbital period of 2~day,
since the gravitational interaction with the parent star is too strong to be
destabilized.

We then tested the possibility of an additional fourth planet in the system
by varying the semi-major axis, the eccentricity, and the longitude of the
periastron over a wide range, and performing a stability analysis (Fig. \ref{F10}). 
The test was completed for a fixed value $K=1$~m/s, corresponding to an Earth-size
object.
We also performed a simulation of a Neptune-size object
($K=10$~m/s) without significant changes in its dynamics.
In this last case, however, an object of this size would have already been
detected in the data.

From this analysis (Fig. \ref{F10}), one can see that stable orbits are possible
beyond 1~AU and, very interestingly, also for orbital periods around 15~days
($\sim$0.083~AU), which correspond to bodies trapped in a 4:1 mean motion resonance
with planet $b$ (and 2:1 resonance with planet $c$).
Among the already known planets, this is the only zone where additional
planetary mass companions can survive.
With the current HARPS precision of $\sim$1~m/s, we estimate that any object
with a minimum mass $M > 2 M_\oplus$ would already be visible in the data.
Since this does not seem to be the case, if we assume that a planet exists in
this resonant stable zone, it should be an %Earth-sized object or smaller.
object smaller or not much larger than the Earth.

The presence of this fourth planet, not only fills an empty gap in the
system, but can also help us to explain the anomalous high eccentricity of planet
$d$ ($e_d \sim 0.14$).
Indeed, tidal interactions with the star should have circularized the orbit in less
than 1~Myr, but according to \citet{Mardling_2007}, the presence of an Earth-sized
outer planet may delay the tidal damping of the eccentricity.

We propose that additional observational efforts should be made to search for this
planet. The orbital period of only 15~day does not require long time of
telescope and a planet with the mass of the Earth is at reach with the present
resolution.
Moreover, assuming coplanar motion with the outermost planets, we obtain
the exact mass for this planet and not a minimal estimation.
We have reanalyzed the residuals of the system (Fig.\,\ref{F3}), but so far no
signal around 15~days appears to be present with the current precision.
However, since this planet will be in a 2:1 mean motion resonance with planet
$c$, we cannot exclude that the signal is hidden in the eccentricity of the
giant planet \citep{Anglada-Escude_etal_2009}.

%________________________________________________________________

\section{Discussion and conclusion}

We have reanalyzed the planetary system orbiting the star GJ\,876, using
the new high-precision observational data acquired with HARPS.
Independently from previous observations with other instruments, we can confirm
the presence of planet $d$, orbiting at 1.93785 days, but in an elliptical orbit
with an eccentricity of 0.14 and a minimum mass of $ 6.3\,M_\oplus$.

By combining the HARPS data with the data previously taken at the
Keck Observatory, we are able to fully characterize the $b$ and $c$
planets. We find $i_b = 48.9^\circ \pm
  1.0^\circ $ and $i_c = 48.1^\circ \pm 2.1^\circ$, which infers for the true
  masses of the planets $M_b = 2.64\,M_{\rm Jup}$ and $ M_c = 0.83\,M_{\rm
  Jup}$, respectively. 
  We hence conclude that the orbits of these two planets are nearly coplanar.
The gravitational interactions between the outer planets and the
innermost planet $d$ may also allow us to determine  its orbit more accurately
in  the near future.
With the current precision of {HARPS} of $\sim$1~m/s for {GJ}\,876, we
expect to detect the true inclination and mass of planet $d$ within some
decades.

A dynamical analysis of this planetary system confirms that planets $b$ and $c$
are locked in a 2:1 mean motion resonance, which ensures stability  over 5~Gyr.
In the nominal solution, the resonant angles $ \theta_b =  2 \lambda_b -
\lambda_c + \varpi_b $ and $ \theta_c =  2 \lambda_b - \lambda_c + \varpi_c $
are in libration around $ 0^\circ $, which means that their periastrons are
also aligned.
This orbital configuration may have been reached by means of the dissipative
process of planet migration during the early stages of the system evolution
\citep[e.g.][]{Crida_etal_2008}.  
By analyzing the proper modes, 
we are able to see that the amplitude  libration of the proper resonant argument 
$ \theta =  2 \lambda_b - \lambda_c + \varpi_2^* $ can be as small as $3.5^\circ$, 
of which only $1.8^\circ$ is related to the libration frequency. 
It is thus remarkable that the libration amplitude is so small, owing to 
the width of the stable resonant island being large.
%, and would thus allow for a much larger libration amplitude. 
We can thus assume that the libration amplitude has been damped by some 
dissipative process. 
This singular planetary system may then provide important constraints on planetary
formation and migration scenarios.

%- presentation d'un modele de fit rapide

Finally, we have found that stability is possible for an Earth-size mass planet
or smaller in an orbit around 15~days, which is in a 4:1 mean motion resonance
with planet $b$.
The presence of this fourth planet, not only fills an empty gap in the
system, but can also help us to explain the anomalous high eccentricity of planet
$d$ ($e_d \sim 0.14$), which should have been damped to zero by tides.
Because of the proximity and low mass of the star, a planet with the mass of the
Earth should be detectable at the present HARPS resolution.
%We then think that an observational effort could be done in searching for this planet. 

We conclude that the radial-velocity technique is self sufficient for fully
characterizing and determining
all the orbital parameters of a multi-planet system, without needing to use
astrometry or transits.

%\vskip0.5cm

%\appendix

%\section{The mean potential energy.}

%\label{ApenA}

%Jocelyn...

\begin{acknowledgements}
We acknowledge support from the Swiss National Research Found (FNRS), the Geneva
University and French CNRS. A.C. also benefited from a grant by the Funda\c{c}\~ao
para a Ci\^encia e a Tecnologia, Portugal (PTDC/CTE-AST/098528/2008). 
\end{acknowledgements}

\bibliographystyle{aa}
\bibliography{correia}

\end{document}